\documentclass[9pt,twoside]{article}

\usepackage{lipsum} 

\usepackage[sc]{mathpazo} 
\usepackage[T1]{fontenc} 
\usepackage{microtype} 

\usepackage[hmarginratio=1:1,top=32mm,columnsep=20pt,a4paper]{geometry} 

\usepackage{multicol} 
\usepackage[small,labelfont=bf,up,textfont=it,up]{caption} 
\usepackage{booktabs} 
\usepackage{float} 
\usepackage{hyperref} 
\usepackage{setspace}

\usepackage{lettrine} 
\usepackage{paralist} 

\usepackage{abstract} 

\usepackage{titlesec} 
\renewcommand\thesection{\Roman{section}} 
\renewcommand\thesubsection{\Roman{subsection}} 
\titleformat{\section}[block]{\large\scshape\centering}{\thesection.}{1em}{} 
\titleformat{\subsection}[block]{\large}{\thesubsection.}{1em}{} 

\usepackage{fancyhdr} 
\usepackage{graphicx}
\usepackage[outdir=./]{epstopdf}
\usepackage{graphics}
\usepackage{amsmath,amssymb,amsfonts,dsfont,amsopn} 
\usepackage{hyperref}
\usepackage{algorithm,algorithmic}
\usepackage[sort,authoryear]{natbib}
\usepackage{epstopdf}

\usepackage{caption}
\usepackage{float}

\usepackage{colortbl}
\usepackage[table]{xcolor}

\definecolor{shadingcolor}{rgb} {.87,    0.92,    .98}
\newcommand{\shadingbox}[1]{   
    \fboxsep 0pt
    \colorbox{shadingcolor}{
        {\hskip -2pt #1}\hskip -2.5pt
    }
}

\newcommand{\minitab}[2][l]{\begin{tabular}{@{}#1}#2\end{tabular}}

\def\x{{\mathbf x}}
\def\y{{\mathbf y}}
\def\w{{\mathbf w}}
\def\v{{\mathbf v}}
\def\u{{\mathbf u}}
\def\a{\mathbf{a}}
\def\f{\mathbf{f}}
\def\c{\mathbf{c}}
\def\btheta{\boldsymbol\theta}

\def\X{\mathbf{X}}

\def\A{\mathbf{A}}

\def\Y{\mathbf{Y}}

\def\D{\mathbf{D}}
\def\S{\mathbf{S}}
\def\0{\mathbf{0}}

\def\Q{\mathbf{Q}}
\def\R{\mathbf{R}}
\def\M{\mathbf{M}}

\def\tX{\underline{\mathbf{X}}}

\def\tG{\underline{\mathbf{G}}}

\def\tY{\underline{\mathbf{Y}}}

\def\t0{\underline{\mathbf{0}}}

\def\tM{\underline{\mathbf{M}}}

\def\PHI{\mathbf{\Phi}}
\def\tPHI{\underline{\mathbf{\Phi}}}

\def\R{\mathds{R}} 

\title{\vspace{-15mm}\fontsize{18pt}{10pt}\selectfont\textbf{Sparse multiway decomposition for analysis and modeling of diffusion imaging and tractography}} 

\author{
\large
\textsc{Cesar F. Caiafa}\thanks{Also with Facultad de Ingenier\'{i}a, University of Buenos Aires, ARGENTINA.}\\[2mm] 
\normalsize Instituto Argentino de Radioastronom\'{i}a (IAR) - CONICET, CCT-La Plata, \\
\normalsize Buenos Aires, ARGENTINA. \\ 
\normalsize \href{mailto:ccaiafa@gmail.com}{ccaiafa@gmail.com}\\[3mm] 
\textsc{Franco Pestilli}\\[2mm] 
\normalsize Dept. of Psychological and Brain Sciences, Programs in Neuroscience and Cognitive Science,\\ 
\normalsize Indiana Network Science Institute, Indiana University, Bloomington, IN, USA.\\ 
\normalsize \href{mailto:franpest@indiana.edu}{franpest@indiana.edu} 
\vspace{-5mm}
}
\date{}

\begin{document}

\maketitle 

\thispagestyle{fancy} 


\begin{abstract}
The number of neuroimaging data sets publicly available is growing at fast rate. The increase in availability and resolution of neuroimaging data requires modern approaches to signal processing for data analysis and results validation. We introduce the application of sparse multiway decomposition methods \citep{Caiafa:2012iv} to linearized neuroimaging models. We show that decomposed models are more compact but as accurate as full models and can be successfully used for fast data analysis. We focus as example on a recent model for the evaluation of white matter connectomes \citep{Pestilli:2014kk}. We show that the multiway decomposed model achieves accuracy  comparable to the full model, while requiring only a small fraction of the memory and compute time. The approach has implications for a majority of neuroimaging methods using linear approximations to measured signals. 


\end{abstract}


\begin{multicols}{2} 

\section{Introduction}
\label{intro}

\begin{figure*}
 \centering
 \includegraphics[width=.9\textwidth]{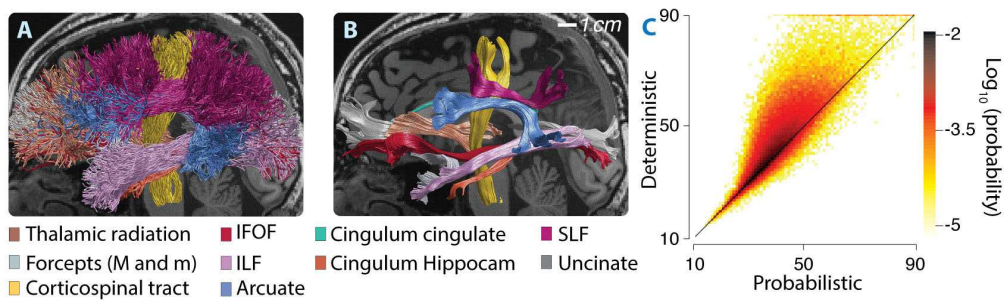}
 \caption{Tractography comparison and evaluation. \textbf{A}. Ten major human white matter tracts generated using probabilistic tractography. \textbf{B}. Ten major human white matter tracts generated using deterministic tractography. \textbf{C}. Comparison of r.m.s. errors in predicting the diffusion signal of probabilistic and deterministic connectomes.}
 \label{fig:Fig_Intro}
\end{figure*}

\lettrine[loversize=-0.4, lines=1, nindent=0em, slope=0em]{N}euroscience is transforming. Investigators have recently started sharing  brain data collected on large human populations \citep{VanEssen:2012bo,VanEssen:2013ep,Calhoun:2011de}. The modern era of data sharing has the potential to promote scientific replicability of results and advance our understanding of brain function. Shared data can be accessed by different research groups and analytic tools can be tested and applied to the very same data in attempt to extend and replicate scientific results \citep{Zuo:2014jr,Pestilli:2015kr}. This fundamental shift of modern neuroscience will allow separating the wheat from the chaff. This paradigm shift is changing the landscape of brain imaging and motivating investigators to implement new methods for data storage, analysis and sharing \citep{Pestilli:2015kr}.  

Hereafter, we show that modern neuroimaging can benefit from recent developments in signal processing \citep{Caiafa:2012iv,Cichocki:2015jja} to address the new challenges raised by the most recent growth in data size, resolution and neuroscience models complexity. Standard signal processing methods in neuroimaging are based on classical Linear Algebra and use mathematical objects such as one dimensional vectors ($1$D) and two dimensional matrices ($2$D).  Multiway approaches exploit the properties of arrays of higher order; arrays with three or more dimensions. Multiway arrays, also referred to as hypermatrices or tensors, are the generalization of vectors and matrices to higher number of dimensions ($N$D).  Multiway arrays have two primary advantages when applied to neuroimaging data compared to classical linear algebra approaches. First, neuroimaging data are multidimensional in nature and their structure is preserved by multiway arrays, which in turns allows for straightforward data addressing. Second, they provide convenient and compact representations of neuroimaging models that allows substantial reduction in memory consumption \citep{Cichocki:2015jja,Morup:2011bh,Caiafa:2013jr}. Below we briefly introduce basic concepts for multyway array decomposition and show one example application of the methods to modeling diffusion imaging and tractography \citep{Pestilli:2014kk}. 
 
dMRI data in combination with fiber tracking allows measuring the anatomy and tissue properties of the human white-matter in living brains \citep{Zhang:2010ju,Yendiki:2011jg,Yeatman:2012ku}.  By measuring living brains this technology allows correlating the properties of the white matter tissue and structure with human behavior, cognition as well as development and aging in health and disease \citep{Thomason:2011ei,SamanezLarkin:2015fa}. dMRI and tractography generate estimates of white matter tracts and connections, such estimates need methods for routine evaluation and validation \citep{Jones:2013gf,2012Sci3371605C,Sporns:2005ki}. For example, it is  established that different tractography methods can generate different estimates of the shape of the white matter anatomy (Fig. \ref{fig:Fig_Intro}A and B). 

Recently we developed a method called LiFE, Linear Fascicle Evaluation \citep{Pestilli:2014kk}. LiFE is an approach to statistical validation of in vivo connectomes that can be applied to individual brains. To validate connectomes in each brain, LiFE requires as input a set of candidate white-matter fascicles generated using available tractography algorithms \citep{Tournier:2012bg,Cook:2006uo,Jiang:2006hs}. The candidate fascicles are used to generate a prediction of the anisotropic diffusion signal measured within the white matter volume using dMRI. The error (root mean squared error, r.m.s.) in predicting the diffusion signal is computed using cross-validation and used to establish the accuracy of a tractography solution. For example, Fig. \ref{fig:Fig_Intro}C shows a comparison of the cross-validated r.m.s. error of two tractography methods in the same brain (the models shown in Fig. \ref{fig:Fig_Intro}A and \ref{fig:Fig_Intro}B). The result demonstrate that in a majority of the white matter the model in Fig. \ref{fig:Fig_Intro}A has lower r.m.s. error in predicting the diffusion signal. 

LiFE predicts the demeaned dMRI data by finding the linear combination of the diffusion prediction contributed by all individual fascicles within a connectome.

\begin{figure}[H]
 \centering
 \includegraphics[width=0.42\textwidth]{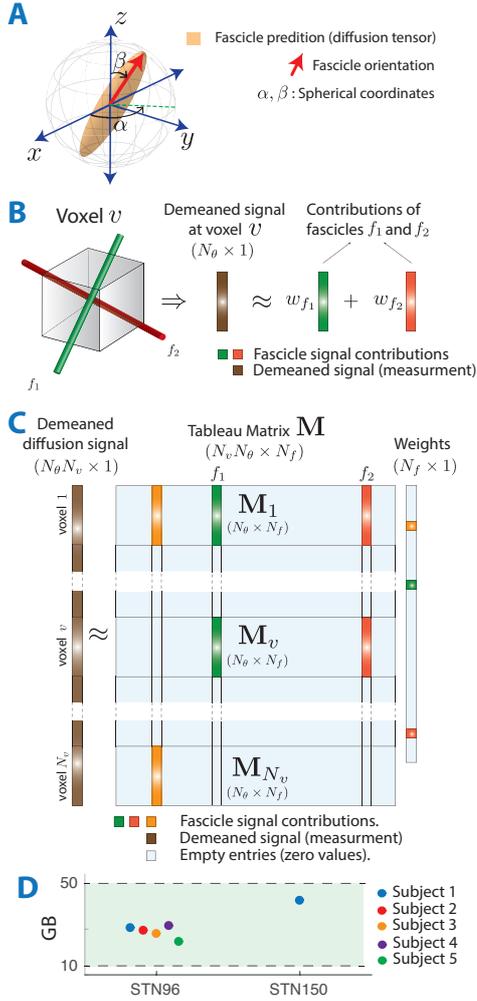}
 \caption{The linear fascicle evaluation method (LiFE). \textbf{A}. Idealized diffusion prediction model for the fascicles in a voxel. The fiber-predicted diffusion signal in a voxel is modeled by the Steijkal-Tanner equation (orange surface; \citet{Stejskal:1965kr,Basser:2011jq}), evaluated in the direction of each fascicle in the voxel (red arrow).  \textbf{B}. The diffusion prediction in a voxel with multiple fascicles is computed as a linear combination of the prediction of each individual fascicle (Eq. \ref{linear_model}). The demeaned diffusion signal (brown) is predicted as the weighted sum of the predictions from all fascicles (green and red) in the voxel. \textbf{C}. Global linear model. The demeaned measured diffusion signal across all white-matter voxels is organized in a long $1$D vector ($\y$). The signal prediction from all fascicle across the full white-matter volume is organized into a large $2$D block-sparse matrix $\M$. Weights are assigned to each fascicle using an optimization procedure, fascicle weights are organized into a $1$D vector ($\w$). Notations, $N_\theta$ represents the number of diffusion directions, $N_v$ the number of voxels, and $N_f$ the number of fascicles). \textbf{D}. LiFE storage measurements.  The size of the LiFE Model ($2$D matrix $\M$ in Panel C) measured in $GB$ in several brains and data sets. Matrices built using double floating-point precision.}
 \label{fig:Fig_LiFE}       
\end{figure}

The fundamental problem solved by LiFE is to allow computing an error measure that can be used for a variety of tasks in evaluating individual connectomes properties generated with virtually any tractography method \citep{Pestilli:2014kk,Gomez:2015fq,Takemura:2015jr,Yeatman:2014gh}. One consequence of formulating the LiFE model as a system of linear equations is the size of the $2$D matrix representing the model (see Fig. \ref{fig:Fig_LiFE}C and Eq. \ref{eq:matrix_equation}). The computational demands of the LiFE method represented as $2$D matrix are substantial. Given an approximate human white-matter volume of $500 ml$, the spatial and directional resolution of modern dMRI measurements and a reasonable number of fascicles in a connectome, the size of the LiFE model can be as large as about $50 GB$ (see Fig. \ref{fig:Fig_LiFE}D). 

Hereafter, we introduce a multiway decomposition method that reduces the size of the LiFE model by over $97\%$. The new multiway approach maintains memory consumption under $1 GB$ per brain. The approach can be applied to any linearized model for global tractography, tractography evaluation and microstructure estimation. Below, we (1) describe the mathematics of the sparse multiway decomposition, (2) show that memory consumption is quasi-constant as function of connectome size when using the decomposition and (3) replicate major results of the original LiFE work using the new approach and several datasets \citep{Pestilli:2014kk}. We provide open source software implementing the LiFE model at {\small \url{github.com/brain-life/life}} and {\small \url{francopestilli.github.io/life}}. Scripts and data used to generate the results in the present articles can be found at these repositories.

\section{Materials and Methods}
\label{Materials}
\label{dWMRI}
\subsection{Diffusion-weighted MRI acquisition}
Diffusion-weighted Magnetic Resonance Imaging data (dMRI) were collected in five males subjects (age 37-39) at the Stanford Center for Cognitive and Neurobiological Imaging using a 3T General Electric Discovery 750 (General Electric Healthcare) equipped with a 32-channel head coil (Nova Medical). This dataset is published \citep{Pestilli:2014kk,Rokem:2015kb} and publicly available at {\small \url{http://purl.stanford.edu/rt034xr8593}} and {\small \url{http://purl.stanford.edu/ng782rw8378}}. Data collection procedures were approved by the Stanford University Institutional Review Board. Written consent was collected from each participant.

\textit{Stanford $96$ diffusion directions data set (STN96):} two diffusion-weighted with whole-brain volume coverage were acquired in five individuals within a single scan session using a dual-spin echo diffusion-weighted sequence. Water-proton diffusion was measured using 96 directions chosen using the electrostatic repulsion algorithm \citep{Jones:1999gu}. Diffusion-weighting gradient strength was set to $2,000 s/mm^{2}$ ($TE$ $= 96.8 ms$). Data were acquired at $1.5 mm^{3}$ isotropic spatial resolution.  Individual data sets were acquired twice and averaged in k-space ($NEX = 2$). Ten non-diffusion-weighted ($b = 0$) images were acquired at the beginning of each scan.

\textit{Stanford $150$ diffusion directions data set (STN150):} for one subject two dMRI data sets were also acquired in a single session using $150$ directions, $2 mm^{3}$ isotropic spatial resolution and b values of $2,000 s/mm^{2}$ ($TE = 83.1, 93.6$, and $106.9 ms$).

MRI images for STN96 and STN150 were corrected for spatial distortions due to B0 field inhomogeneity. To do so, B0 magnetic field maps were collected with the same slice prescription as the dMRI data using a 16-shot, gradient-echo spiral-trajectory pulse sequence. Two volumes were acquired before and after the dMRI scan ($TE =9.091ms$ and $11.363 ms$ respectively), the phase difference between the volumes was used as an estimate of the magnetic field.

Subjects' motion was corrected using a rigid-body alignment algorithm \citep{Friston:2004hs}. Diffusion gradients were adjusted to account for the rotation applied to the measurements during motion correction. The dual-spin echo sequence we used does not require eddy current correction because it has a relatively long delay between the RF excitation pulse and image acquisition. This allows for sufficient time for the eddy currents to dephase. Processing software is available at {\small \url{https://github.com/vistalab/vistasoft}}.

\subsection{Anatomical MRI acquisition and tissue segmentation}
The white/gray matter border was defined on the average of two 0.7-mm3 T1-weighted FSPGR images acquired in the same scan session. Tissue segmentation was performed using an automated procedure (FreeSurfer \citet{Fischl:2012el}) and refined manually ({\small \url{http://www.itksnap.org/pmwiki/pmwiki.php}}).

\subsection{Whole-brain connectomes generation and visualization}
Fiber tracking was performed using the MRtrix 0.2 toolbox \citep{Tournier:2012bg}.  White-matter tissue was identified from the cortical segmentation performed on the T1-weighted images and resampled at the resolution of the dMRI data. Only white-matter voxels were used to seed fiber tracking. We used two tracking methods: (i) tensor-based deterministic tractography \citep{Tournier:2012bg,Basser:2000hl,Lazar:2003ds} and (ii) CSD-based probabilistic tracking \citep{Tournier:2012bg,Behrens:2003un,Parker:2003hs} with a maximum harmonic order of $10$ ($L_{max} = 10$, step size: $0.2 mm$; minimum radius of curvature, $1 mm$; maximum length, $200 mm$; minimum length, $10 mm$; fibers orientation distribution function ($f_{ODF}$) amplitude cutoff,  was set to $0.1$).

We created candidate whole-brain connectomes with $500,000$ fascicles in each individual brain (five), data set (two) and tractography method (two). Analysis were performed independently for each brain. Figures of tracts and brain images were generated using the Matlab Brain Anatomy toolbox: {\small \url{https://github.com/francopestilli/mba}}.

\subsection{Brief introduction to modeling magnetic resonance diffusion signals from living human brain tissue}
\label{Diffusion}

The brain tissue comprises different cell types (e.g., neurons, astrocytes and oligodendrocytes). dMRI measures signals that depends on the combination of all cellular components within the brain tissue. The dMRI measurements are generally modeled as the linear combination of two primary components.  One component describes the directional diffusion signal and is presumably related primarily to the direction of the neuronal axons wrapped in myelin sheaths (white matter). This signal is often referred to as \textit{anisotropic diffusion}. The other component describes \textit{isotropic diffusion} (non-directional) and is presumably related to the combination of signals originating from the rest of the cellular bodies within the brain tissue. Below we describe the simplest, fundamental equations used to model the measured dMRI signal in relation to the brain tissue components.

dMRI measures the diffusion signal with and without diffusion sensitization. The measured signal results from the combination of diffusion gradient strength, duration and depends on the interval between two applied gradient pulses. Below we denote the strength of sensitization with $b$ and the vectors of measured diffusion directions (directions along which the diffusion-sensitization gradients are applied) using the unit-norm vector $\btheta\in{\R^3}$.

For a given sensitization strength $b$ and diffusion directions $\btheta$, the diffusion signal measured at each location within a brain (voxel $v$) can be computed using the following equation \citep{Frank:2002bs,Behrens:2003kf}:

\begin{equation} \label{diffusion_model} \small
S(\theta,v) \approx \w_0 S_0(v) e^{-A_0} + \sum_{f\in v} \w_fS_0(v)e^{-b\btheta^T\Q_f \btheta},
\end{equation}
where $f$ is the index of the candidate white-matter fascicles within the voxel, $S_0(v)$ is the non diffusion-weighted signal in voxel $v$ and $A_0$  is the isotropic apparent diffusion (diffusion in all directions). The value $\btheta^T\Q_f \btheta > 0$ gives us the apparent diffusion at direction $\btheta$ generated by fascicle $f$. $\Q_f\in {\R^{3\times 3}}$ is a symmetric and positive-definite matrix called tensor \citep{Basser:1994ir}.  The tensor allows a compact representation of the diffusion signal measured with dMRI. For example, $\Q_f$ in the exponent of Eq. \ref{diffusion_model} can be replaced with the following simple tensor model:
\begin{equation}\small
\Q_f = 
\begin{array}{ccc}
[
\u_1 & \u_2 & \u_3 
]
\end{array} 
\left[
\begin{array}{ccc}
s_a & 0 & 0 \\
0 & s_{r_1} & 0 \\
0 & 0 & s_{r_2} 
\end{array} 
\right]
\left[
\begin{array}{c}
\u_1 \\
\u_2 \\
\u_3 
\end{array}
\right], 
\end{equation}
where $\u_n \in {\R^{3\times 1}}$ are the unit-norm orthogonal vectors that correspond to the semi-axes of the diffusion tensor ellipsoid, and $s_a$, $s_{r_1}$, $s_{r_2}$ define the axial and radial diffusivity of the tensor, respectively. In the simplest version of the fascicle model, $s_a=1$ and $s_{r_1}=s_{r_2}=0$, which means that diffusion is restricted to the main axis direction \citep{Behrens:2003kf,Pestilli:2014kk}.

\subsection{Brief introduction to multiway arrays}
\label{Notation}

Multiway arrays generalize vectors ($1$D array) and matrices
($2$D array) to arrays of higher dimensions, three or more. Such arrays can be used to perform multidimensional factor analysis and decomposition and are of interest to many scientific disciplines \citep{Kolda:2009vq,Cichocki:2015jja}. Below we introduce a few basic concepts and notation helpful in discussing multiway arrays and their decomposition. We refer the reader to Table \ref{tab:Notation} for a summary of basic notations and definitions.

{\it Multiway arrays.} Vectors ($1$D arrays) and matrices ($2$D arrays) are denoted below using boldface lower- and upper-case letters, respectively. For example  $\x \in{\R^I}$ and $\X\in{\R^{I\times J}}$ represent a vector and a matrix, respectively. A multiway array, also called $N$-th order array, is the generalization of matrices to more than two dimensions and is denoted by an underlined boldface capital letter, e.g. $\tX\in{\R^{I\times J \times K}}$ is an $3$-rd order array of real numbers. Elements $(i,j,k)$ of a multiway array are referred to as $x_{ijk}$. 

\begin{figure}[H]
 \centering
 \includegraphics[width=0.45\textwidth]{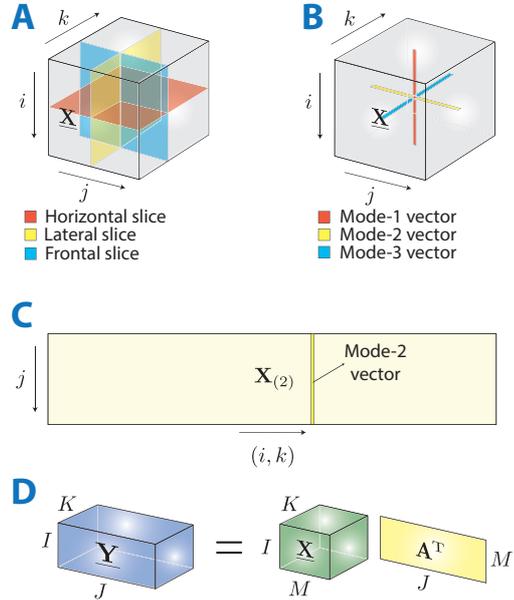}
 \caption{Multiway array decomposition basic notation and operations. \textbf{A}. Frontal, lateral and horizontal slices of an example multiway array. \textbf{B}. Examples of mode-n vectors.  \textbf{C}. Illustration of the mode-2 unfolding matrix $\X_{(2)}$. \textbf{D}. Array-by-matrix product (example product in mode-2). See Table \ref{tab:Notation} for additional information about notation and mathematical definitions.}
 \label{fig:Fig_Nway}       
\end{figure}

{\it Array slices.} Array slices are used to address a multiway array along a single dimension (a cut through a single dimension of the array). Slices are obtained by fixing the index of one dimension of the multiway array while letting the other indices vary. For example, in a $3$-rd order array $\tX\in{\R^{I\times J \times K}}$, we can identify horizontal ($i$), lateral  ($j$) and frontal ($k$) slices by holding fixed the corresponding index of each array dimension (see Fig. \ref{fig:Fig_Nway}A).

{\it Unfolding arrays and mode-$n$ vectors.} Multiway arrays can be conveniently addressed in any dimension by means of mode-$n$ vectors. Given a multiway array $\tX\in{\R^{I\times J \times K}}$, its mode-$n$ vectors are obtained by holding all indices fixed except one, thus corresponding to columns ($n=1$), rows ($n=2$) and so on (see Fig. \ref{fig:Fig_Nway}B). For example, a $3$D multiway array can be converted into a matrix by re-arranging its entries (unfolding). The mode-$n$ unfolded matrix, denoted by $\X_{(n)}\in{\R^{I_n\times \bar{I}_n}}$, where $\bar{I}_n=\prod_{m\ne n} I_m$ and whose entry at row $i_n$ and column $(i_1 -1) I_2 \cdots I_{n-1} I_{n+1} \cdots I_N +  \cdots + (i_{N-1} -1)I_N + i_N$ is equal to $x_{i_1 i_2 \ldots i_N}$. For example, mode-$2$ unfolding builds the matrix $\X_{(2)}$ where its columns are the mode-$2$ vectors of the multiway array and the rows are vectorized versions of the lateral slices, i.e. spanning dimensions with indices $i$ and $k$ (see Fig. \ref{fig:Fig_Nway}C). 

{\it Multiway array by matrix product.} Multiway arrays can be multiplied by matrices ($2$D arrays) only if the matrix and multiway array sizes match in the mode specified for the multiplication. This is a generalization of matrix multiplication. Given a multiway array $\tX\in{\R^{I_1\times I_2\cdots \times I_N}}$ and a matrix $\A \in{\R^{J\times I_n}}$, the mode-$n$ product 
\begin{equation}
\tY=\tX \times_n \A \in{\R^{I_1\times \cdots \times I_{n-1}\times J \times I_{n+1}\cdots \times I_N}}
\end{equation}
is defined by: 
\begin{equation}\label{tenbymat_prod}
y_{i_1 \cdots i_{n-1} j i_{n+1} \cdots i_N} = \sum_{i_n=1}^{I_n} x_{i_1\cdots i_n \cdots i_N}a_{ji_n},
\end{equation}
with $i_k=1,2,...,I_k$ ($k\neq n$) and $j=1,2,...,J$. It is noted that this operation involves the products of matrix $\A$ by each one of the mode-$n$ vectors of $\tX$. In Fig. \ref{fig:Fig_Nway}B, the $3$-rd order array by matrix product in mode-$2$ is illustrated, i.e. $\tY = \tX\times_2 \A$. The size of the resulting array $\tY$ is that of $\tX$ in mode-$1$ and -$3$ while the size of $\tY$ in mode-$2$ is $J$, i.e. it is equal to the number of rows in $\A$. 

\begin{table*}
\caption{Mathematical notation and definitions for multiway arrays and decomposition.} \centering
 {\shadingbox{\scriptsize
 \begin{tabular*}{\linewidth}[t]{@{\extracolsep{\fill}}ll} \hline \\[-2ex]
 {$\tX, \; \A, \; \w, \; b$}& {A tensor, a matrix, a vector and a scalar} \\[0.2cm]
 {$x_{i_1i_2\dots i_N}, \; a_{ij}, \; w_i$}& {Entries of a tensor, a matrix and a vector} \\[0.2cm]
 {$\tX(:,j,k), \; \tX(i,:,k), \; \tX(i,j,:)$}& {Mode-1, mode-2 and mode-3 vectors are obtained by fixing all but one index} \\[0.2cm]
 {$\tX(i,:,:), \; \tX(:,j,:), \; \tX(:,:,k)$}& {Horizontal, lateral and frontal slices are obtained by fixing all but two indices} \\[0.2cm]
 $\X_{(n)} \in \R^{I_n \times I_1 I_2 \cdots I_{n-1} I_{n+1} \cdots I_N}$ & \minitab[p{.62\linewidth}]{Mode-$n$ unfolding of multiway array $\tX \in \R^{I_1 \times I_2 \times \cdots \times I_N}$ whose entry at row $i_n$ and column $(i_1 -1) I_2 \cdots I_{n-1} I_{n+1} \cdots I_N +  \cdots + (i_{N-1} -1)I_N + i_N$ is equal to $x_{i_1 i_2 \ldots i_N}$}\\[0.5cm]
$\tY=\tX \times_n \A \in{\R^{I_1\times \cdots \times I_{n-1}\times J \times I_{n+1}\cdots \times I_N}}$ & \minitab[p{.62\linewidth}]{Multiway array by matrix product (in mode-$n$) where  $y_{i_1 \cdots i_{n-1} j i_{n+1} \cdots i_N} = \sum_{i_n=1}^{I_n} x_{i_1\cdots i_n \cdots i_N}a_{ji_n}$}\\[0.4cm]
{$\tX  \approx \tG \times_1 \A_1 \times_2 \A_2 \times_3 \A_3$}& \minitab[p{.62\linewidth}]{Tucker decomposition: a 3D multiway array $\tX\in{\R^{I_1\times I_2\times I_3}}$ is represented as the product of a core array $\tG\in{\R^{R_1\times R_2\times R_3}}$ by factor matrices $\A_n\in{\R^{I_n\times R_n}}$}\\[0.4cm]
{$\x = vec(\tX)\in{\R^{I_1I_2\cdots I_N}}$}& \minitab[p{.62\linewidth}]{Vectorization of multiway array $\tX\in{\R^{I_1\times I_2\times \cdots \times I_N}}$ with the entry at position $i_1+\sum_{k=2}^N[(i_k-1)I_1I_2\cdots I_{k-1}]$ equal to $x_{i_1i_2\cdots i_N}$ }\\[0.4cm]
{$\S_0 = diag(S_0(1),S_0(2),\dots,S_0(N_v))$}& {Diagonal $N_v\times N_v$ matrix with $\S_0(v,v) = S_0(v)$} \\[0.2cm]
\hline
\end{tabular*}
} }
\label{tab:Notation}
\end{table*} 

\section{Results}
\label{res}

\subsection{The Linear Fascicle Evaluation model}
\label{MatrixLiFE}

The Linear Fascicle Evaluation method (LiFE; \citet{Pestilli:2014kk}) allows evaluating the accuracy of connectomes generated using dMRI and any fiber tracking algorithm.  The method evaluates the tractography solution by estimating the contribution to predicting the measured diffusion signal from all fascicles contained in a connectome (see Fig. \ref{fig:Fig_LiFE}A and B for examples of fascicles). The evaluation method focuses on the fascicles, for this reason it predicts the anisotropic diffusion signal; the demeaned diffusion signal (signal that is independent of isotropic diffusion, first term in the right hand of Eq. \ref{diffusion_model}). The anisotropic signal within a voxel $v$ is predicted by demeaning the signal prediction in the following way:
\begin{equation}\label{linear_model}
S(\theta,v) - I_v \approx \sum_{f\in v} w_f O_f(\theta), 
\end{equation}
where the mean $I_v$ is defined as
\begin{equation}
I_v = \frac{1}{N_\theta}\sum_\theta S(\theta,v),
\end{equation}
with $N_\theta$ being the number of sensitization directions and $O_f(\theta)$ is orientation distribution function specific to each fascicle, i.e. the anisotropic modulation of the diffusion signal around its mean and it is defined as follows:
\begin{equation}\small
\label{eq:o_f}
O_f(\theta) = S_0(v)\left( e^{-b\btheta^T\Q_f \btheta} - \frac{1}{N_\theta}\sum_\theta e^{-b\btheta^T\Q_f \btheta} \right).
\end{equation}
The left-side hand in Eq. \ref{linear_model}, the demeaned signal, is the difference between the measured diffusion signal and the
mean diffusion signal. The right-hand side of Eq. \ref{linear_model} is the prediction model. The LiFE model extends from the single voxel to all white-matter voxels in the following way:

\begin{equation}\label{eq:matrix_equation}
\y \approx \M \w,
\end{equation} 
where $\y\in{\R^{N_\theta N_v}}$ is a vector containing the demeaned signal for all white-matter voxels $v$ and across all diffusion directions $\theta$, i.e. $y_i = S(\theta_i,v_i) - I_{v_i}$. The matrix $\M \in{\R^{N_\theta N_v \times N_f}}$ contains at column $f$ the signal contribution $O_f(\theta)$ given by fascicle $f$ at voxel $v$ across all directions $\theta$, i.e., $\mathbf{M}(i,f) = O_f(\theta_i)$, and $\w\in{\R^{N_f}}$ contains the weights for each fascicle in the connectome. These weights are estimated by solving the following non-negative least-square constrained optimization problem:
\begin{equation}
\label{nnls}
\min_{\mathbf{w}}\left( \frac{1}{2}\|\mathbf{y} - \mathbf{M}\mathbf{w}\|^2\right)  \mbox{   subject to } w_f \ge 0, \forall f.
\end{equation}

This formulation of the LiFE model requires generating a matrix $\M$ (Fig. \ref{fig:Fig_LiFE}C) that is very large and block-sparse. $\M$ is sparse because fascicles only cross a subset of all voxels. The matrix has $N_\theta N_v$ rows by $N_f$ columns. Given an approximate human white-matter volume of 500 $ml$ the number of voxels ($N_v$) can vary between $50,000$ and $500,000$ depending on the spatial resolution of the dMRI acquisition (normally between 4-1.25 $mm^3$). Given that modern dMRI acquisition parameters measure between $30$ and $300$ diffusion directions ($N_\theta$) and that whole-brain connectomes can contain as little as $100,000$ but as much as $10,000,000$ fascicles (columns of $\M$), the size of $\M$ can vary between $10$ and $100$GB using standard double-precision floating-point format and sparse format. 

Fig. \ref{fig:Fig_LiFE}D shows measurements of the size of $\M$ in $GB$ for five individual brains and two datasets. Such large memory requirements necessitate large-memory compute-systems nowadays available in major research institutions (e.g., {\small \url{http://rt.uits.iu.edu/bigred2/}}, {\small \url{http://karst.uits.iu.edu}}). Below we introduce a novel approach that dramatically reduces the storage requirements of the LiFE model and makes LiFE suitable to run on standard desktop and notebook computers.

\subsection{Multiway decomposition of the Linear Fascicle Evaluation model}
\label{naturalmultiway}
Major contribution of the present work is a method to represent the LiFE model via Sparse Tucker Decomposition (STD; \citep{Caiafa:2012iv}). Before introducing the method, we briefly discuss the standard and sparse Tucker-decomposition methods (see also Methods \ref{tab:Notation}).

\begin{figure}[H]
 \centering
 \includegraphics[width=0.45\textwidth]{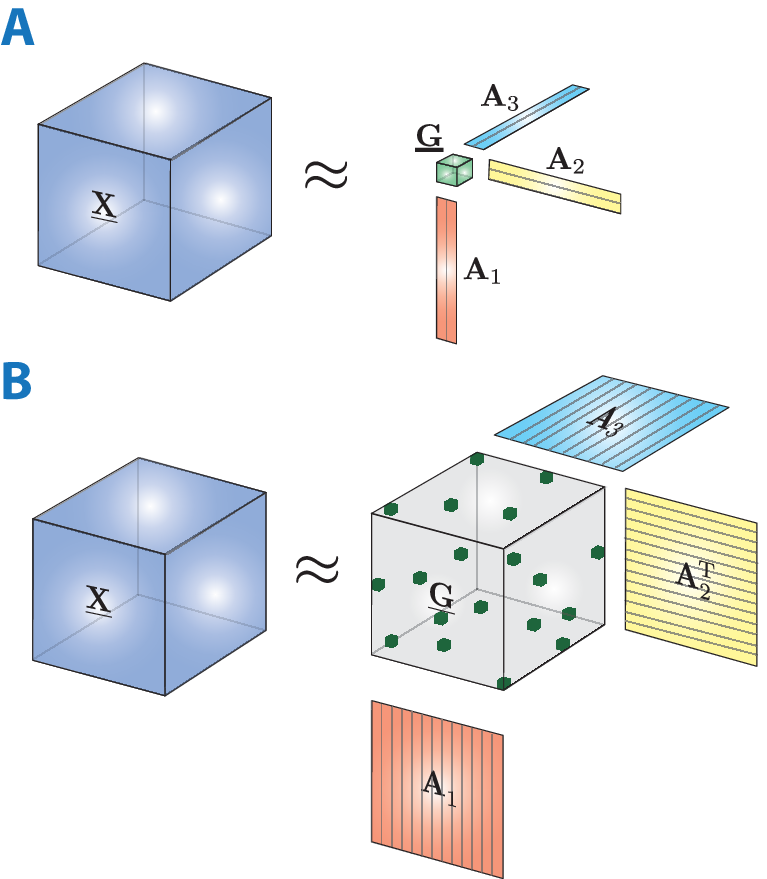}
 \caption{Classical and sparse Tucker decomposition. \textbf{A}. The classical Tucker decomposition \citep{Tucker1966} allows representing a $3$D multiway array $\tX\in{\R^{I_1\times I_2\times I_3}}$ as the product of core array (green) $\tG\in{\R^{R_1\times R_2\times R_3}}$ by factor matrices $\A_n\in{\R^{I_n\times R_n}}$ (red, yellow and blue). Data compression is achieved by considering very small (dense) core arrays $\tG$, meaning that $R_n\ll I_n$. \textbf{B}. The sparse Tucker Decomposition (STD; \citet{Caiafa:2012iv}). The core array $\tG$ is large but sparse, not dense as in the classical Tucker model. Data compression is achieved because of the sparsity of the core array.}
 \label{fig:Fig_Tucker}       
\end{figure}

\textit{Data compression by Tucker decomposition.} Ledyard Tucker, a mathematician who specialized in statistics and psychometrics, was first in proposing a decomposition approach to multidimensional factorization problems \citep{Tucker1966}. The approach provides a generalization of the low-rank approximation for matrices to multiway arrays. Using the Tucker model, a $3$-rd array $\tX\in{\R^{I_1\times I_2\times I_3}}$, is approximated by the following decomposition:
\begin{equation}\label{eq:Tucker}
\tX  \approx \tG \times_1 \A_1 \times_2 \A_2 \times_3 \A_3,
\end{equation}
with a \emph{core array} $\tG\in{\R^{R_1\times  R_2 \times R_3}}$ and \emph{factor matrices} $\A_n \in{\R^{I_n\times R_n}}$. The decomposition is not guaranteed to provide a good aproximation, but when it does it compresses data because it uses a core array that is much smaller than the original multiway array, i.e. $R_n\ll I_n$. Indeed, the Tucker model provides us with a powerful compression method. This is because, instead of storing the whole original multiway array, we can store core array and factors. In this case, Eq. \ref{eq:Tucker} \citep{Tucker1966} is called a \textit{low-rank} Tucker model because $\tG$ is small compared to $\tX$ (see Fig. \ref{fig:Fig_Tucker}A). 

A simple example can help us explain how compression is obtained with a standard low-rank Tucker model. Let consider a situation where $R=R_1=R_2=R_3$ and $I=I_1=I_2=I_3$. The low-rank Tucker decomposition of $\tX$ requires storing only $R^3+3IR$ values. Where $R^3$ is the number of entries of $\tG$ and $I$ and $R$ the dimensions of $\A_1$, $\A_2$ and $\A_3$. Instead, the full multiway array would require storing $I^3$ values, compression is noticeable especially when $\tG$ is small meaning that $R\ll I$.

\textit{Data compression by sparse Tucker decomposition:} If the core  $\tG$ of a Tucker decomposition of a multiway array is sparse, compression can be achieved even when $\tG$ is very large \citep{Caiafa:2012iv}. Consider a Tucker model with a large but sparse core array  $\tG\in{\R^{R_1\times  R_2 \times R_3}}$, see Fig. \ref{fig:Fig_Nway}E. Only some entries in $\tG$ are nonzero, i.e. $g_{i_m,j_m,k_m}\neq 0$ for $m=1,2,\dots M$ where $M$ is the number of nonzero entries. Storing $\tG$ all requires storing: (1) the non-zero coefficients, (2) their location in $\tG$, and (3) the factor matrices $\A_n$. This means that the storage cost (order) of the model is $4M + 3IR$ (assuming that $R=R_1=R_2=R_3$ and $I=I_1=I_2=I_3$). Hereafter, we refer to this model as Sparse Tucker Decomposition (STD). Compared to the classical low-rank Tucker model, STD can provide better compression ratios with relatively low $M$, i.e., with very sparse core arrays $\tG$ \citep{Caiafa:2013jr}. 

Below we show how to apply the STD approach to the LiFE model. To do so we first explain how the diffusion signals within a voxel predicted by the LiFE model can be represented using dictionaries of precomputed diffusion prediction signals. After that we show how to extend the decomposition approach to the whole white matter volume. 

\subsubsection{Decomposing LiFE within a voxel using prediction dictionaries}
\label{sec:dictionary}
\begin{figure*}[t]
 \centering
 \includegraphics[width=1\textwidth]{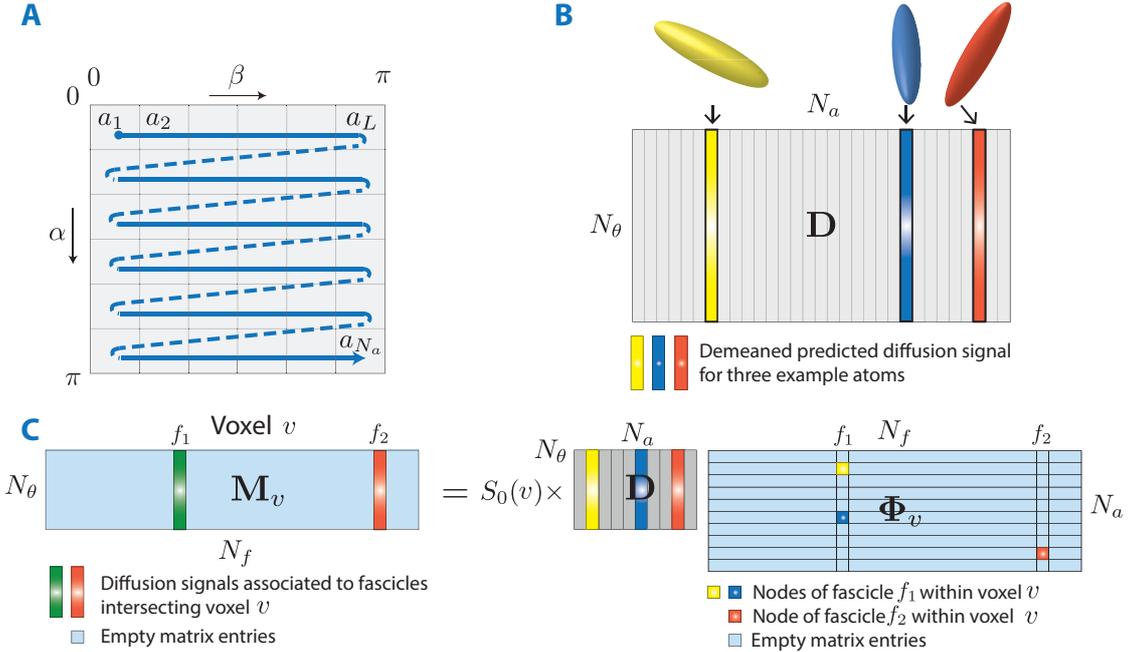}
 \caption{Discretization of the LiFE model. \textbf{A}. Discretization of the unit-norm sphere and mapping of the spherical coordinates to a single index $a$. The unit sphere of diffusion directions is sampled by using a uniform grid of spherical coordinates (using a selected choice of $\alpha$ and $\beta$ values see Fig. \ref{fig:Fig_LiFE}A for definitions). \textbf{B}. Construction of the diffusion prediction dictionary. Demeaned diffusion predictions are generated for each spherical coordinate (orientation) and measured diffusion direction. Predictions are stored in the columns of a dictionary of diffusion predictions, $\D$, whose columns identify different fascicles orientations ($N_a$) and rows the predicted diffusion in a measured diffusion direction ($N_{\theta}$). \textbf{C}. Modeling the diffusion signal in a voxel using the discretized LiFE model. The measured signal in voxel $v$ is organized as a matrix, $M_{v}$. The signal predicted by each fascicle (two in our example) and nodes (three in our example) passing in $v$ is approximated by combining the diffusion prediction of the dictionary atoms (columns of $\D$) with directions closest to the orientation of the fascicles nodes (yellow, blue and red). Non-zero entries in $\PHI_v$ indicate the atoms corresponding to the nodes in the fascicles, $f_{1}$ and $f_{2}$ in the example.
}
\label{fig:Fig_Dictionary}       
\end{figure*}

The original LiFE model predicts anisotropic diffusion within a voxel, $v$, at each measured diffusion direction, $\btheta$, by using the orientation of the white matter fascicles intersecting the voxel (Fig. \ref{fig:Fig_LiFE}B;  \citet{Pestilli:2014kk}). Fascicles are defined as a list of $(x,y,z)$ spatial coordinates within the brain, the fascicle {\it nodes}. To generate the prediction, a demeaned tensor model is evaluated parallel to the orientation of each fascicle node (Eq. \ref{eq:o_f}). Hereafter, we simplify the calculations by representing the signal prediction for any arbitrary fascicle-node orientation using a dictionary of diffusion prediction signals generated for a predefined set of orientations regularly sampled over a grid. This grid covers a plausible rage of fascicles orientations on the unit-norm sphere (shown in Fig. \ref{fig:Fig_LiFE}A) given the resolution of the dMRI data (Fig. \ref{fig:Fig_Dictionary}A). 

We define the dictionary matrix $\D \in{\R^{N_\theta \times N_a}}$ (Fig. \ref{fig:Fig_Dictionary}B) containing in its columns the demeaned canonical diffusion signals called {\it atoms} that can be used to approximate fascicles' contributions. More specifically, we define
\begin{equation}\label{blockM}
\D(\theta,a) = e^{-b\btheta^T\Q_a \btheta} - \frac{1}{N_\theta}\sum_\theta e^{-b\btheta^T\Q_a \btheta}, 
\end{equation}
where $\Q_a$ is the diffusion prediction associated to the atom $a$, i.e. with a single orientation in 3D space. Dictionary atoms are identified by discretizing each spherical coordinate $\alpha$ and $\beta$ using a grid of values in the rage $[0, \pi)$ (Fig. \ref{fig:Fig_Dictionary}B). The parameter $L$ indicates the number of discretization samples on the grid. For example, $L=180$ indicates a grid resolution of $1^\circ \times 1^\circ$ with an approximate atom number $N_a \approx L^2$.

The dictionary $\D \in{\R^{N_\theta \times N_a}}$ allows predicting the matrix $\M_v\in{\R^{N_\theta \times N_f}}$, defined as a block of matrix $\M$ corresponding to voxel $v$ (see Fig. \ref{fig:Fig_Dictionary}C), by decomposing it in the following way:
\begin{equation}\label{eq:dictMvox}
\M_v = S_0(v)\D \PHI_v,
\end{equation}
where $\PHI_v\in{\R^{N_a\times N_f}}$ is a sparse matrix whose non-zero entries at column $f$ indicates the dictionary atoms selected to predict the voxel signal, given the orientations of the fascicles in the voxel. In sum, instead of storing the individual contribution of each fascicle within a voxel, we store only the indices to the dictionary atoms, and avoid multiple versions of very similar signals. Below we extend the STD of LiFE from single voxels to the entire white matter volume. We show that for each data set used we can find a finite number of dictionary atoms that generate predictions as accurate as those obtained with the original LiFE model.

\subsubsection{Extending the LiFE decomposition model across all voxels}
The original LiFE model comprises a single very large block-sparse matrix ($\M$ in Eq. \ref{eq:matrix_equation}). The matrix $\M\in{\R^{N_\theta N_v \times N_f}}$ can be converted into a multiway array by mapping the diffusion directions ($\theta$), voxels ($v$) and fascicles ($f$), onto the dimensions of a 3D multiway array $\tM\in{\R^{N_\theta \times N_v\times N_f}}$.  Representing $\M$ as a 3D multiway array $\tM$ allows us using a Sparse Tucker decomposition approach and compress its size \citep{Caiafa:2012iv}.

\begin{figure}[H]
 \centering
 \includegraphics[width=0.35\textwidth]{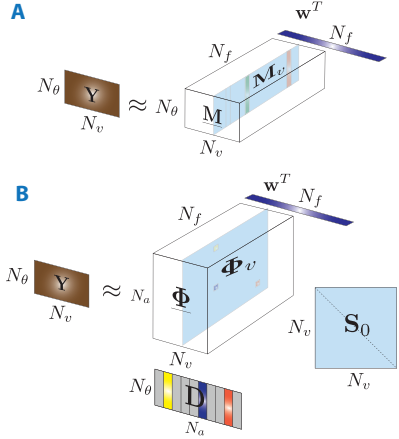}
 \caption{Sparse Tucker Decomposition of LiFE problem. \textbf{A}. Multiway version of the LiFE method $\y \approx \M\w$, Eq. \ref{eq:M_Decomposition}. \textbf{B}. LiFE problem decomposition by Sparse Tucker method, Eq. \ref{eq:LiFEBiD_model}.
}
 \label{fig:Fig_TensorLife}
\end{figure}

To fully exploit the Sparse Tucker Decomposition we consider both the LiFE model $\M$ and its optimization problem, as defined in Eq. \ref{eq:matrix_equation}. This equation can be conveniently rewritten using multiway arrays in the following manner (see Fig. \ref{fig:Fig_TensorLife}A):
\begin{equation}\label{eq:Nway_model}
\Y \approx \tM \times_3 \w^T,
\end{equation}
where matrix $\Y \in{\R^{N_\theta \times N_v}}$ is the matrix version of vector $\y$, and $\tM \in{\R^{N_\theta \times N_v \times N_f}}$ is a 3-way array. The lateral slices of $\tM$ are defined by the matrices $\M_v \in{\R^{N_\theta \times N_f}}$ defined in Eq. \ref{eq:dictMvox}, these represent matrices $\M_{v}$ corresponding to each white matter voxel. Furthermore, Eq. \ref{eq:dictMvox} shows that each lateral slice of $\tM$ is scaled by $S_0(v)$, which we represent at a diagonal matrix $\S_{0}$. This allows us writing the LiFE model that encompasses all voxels by using an efficient Sparse Tucker Decomposition (STD):
\begin{equation}\label{eq:M_Decomposition}
\tM = \tPHI \times_1 \D \times_2 \S_0,
\end{equation}
where the 3-way array $\tPHI \in{\R^{N_a \times N_v \times N_f}}$ has as lateral slices the matrices $\PHI_v \in{\R^{N_a \times N_f}}$ and matrix $\S_0=diag(S_0(1),S_0(2),\dots,S_0(N_v))\in{\R^{N_v \times N_v}}$ is a diagonal matrix with values $S_0(v)$ along the main diagonal. By combining Eqs. (\ref{eq:Nway_model}) and (\ref{eq:M_Decomposition}) we obtain the following Sparse Tucker decomposition of the LiFE model (see also Fig. \ref{fig:Fig_TensorLife}B):
\begin{equation} \label{eq:LiFEBiD_model}
\Y \approx \tPHI \times_1 \D \times_2 \S_0 \times_3 \w^T.
\end{equation}
$\tPHI$ is the core of the multiway decomposition, because $\tPHI$ is sparse it results in strong data compression (see section \ref{sec:sto_red}).
Below we introduce the set of operations necessary to build and optimize (fit) the STD LiFE model.

\subsection{Building the STD LiFE model}
The STD LiFE model is built using a large collection of white matter fascicles estimated using computational tractography, the connectome \citep{Sporns:2005ki}. Fascicles in the connectome are represented as a list of $(x,y,z)$ brain coordinates. The $\tPHI$ core-array of STD LiFE model is then built in the following way: (1) the orientation of each fascicle's ($f$) node ($n$) is identified in spherical coordinates $(\alpha, \beta)$; (2) nodes' orientation is mapped to the closest atom ($a$) in the dictionary ($\D$); (3) the entry in $\tPHI$ corresponding to each identified voxel, fascicle and atom is set to $1$:

\begin{equation}
\tPHI(a,v,f) = 1,
\end{equation}
The rest of the entries in $\tPHI$ are set to zero.

\subsection{Optimizing the STD LiFE model}
Building the STD LiFE model is the first of two steps in evaluating a brain connectome. The final step requires finding the non-negative weights that least-square fit the measured diffusion data. This is a convex problem that can be solved using a variety of Non-Negative Least Squares (NNLS) optimization algorithms (\ref{nnls}). The original LiFE problem, was solved using a NNLS algorithm based on first-order methods \citep{Kim:2013kqa}. Hereafter, we show how to modify the optimization algorithm to exploit the STD LiFE model.

The optimization of the STD LiFE needs to be performed using its core array ($\tPHI$) and matrices ($\D$ and $\S_0$; Eq. \ref{eq:M_Decomposition}). The gradient of the original objective function for the LiFE model can be written as follows: 
\begin{equation}\label{grad}
\nabla_{\w}\left( \frac{1}{2}\|\y - \M\w\|^2\right)  = \M^T \M \w,
\end{equation}
where $\M\in{\R^{N_{\theta}N_v \times N_f}}$ is the original LiFE model, $\w\in{\R^N_f}$ the fascicle weights and $\y\in{\R^{N_{\theta}N_v}}$ the demeaned diffusion signal. Because the STD version does not explicitly store $\M$ in the following section we describe how to perform two basic operations ($\y=\M\w$ and $\w=\M^T\y$) using the multiway decomposition to compute the optimization gradient. Appendix A reports pseudocode implementing the operations.

\subsubsection{Computing $\y = \M\w$}
The product $\M\w$ can be computed in the following way using a 3D-array by vector product:
\begin{equation}
\Y = \tM \times_3 \w^T,
\end{equation}
where the result is a matrix $\Y\in{\R^{N_{\theta}\times N_v}}$, a matrix version of the vector $\y$. Using the STD LiFE model (Eq. \ref{eq:M_Decomposition}) the product is written as follows:

\begin{equation}
\Y = \tPHI \times_1 \D \times_2 \S_0 \times_3 \w^T.
\end{equation}

\subsubsection{Computing $\w = \M^T\y$} 
The product $\w = \M^T\y$ can be computed using the STD LiFE model in the following way \citep{Kolda:2009vq}:
\begin{equation}\label{eq:kron}
\w = \M^T\y = \M_{(3)}\y = \PHI_{(3)}(\S_0 \otimes \D^T)\y,
\end{equation}
where $\otimes$ is the Kronecker product. Eq. \ref{eq:kron} equation can be written as follows:
\begin{equation}\label{eq:Mtranspy}
\w = \PHI_{(3)}vec{(\D^T\Y\S_0)},
\end{equation}
where $vec()$ stands for the vectorization operation, i.e. to convert a matrix to a vector by stacking its columns in a long vector \citep{Caiafa:2012iv}.

Because matrix $\PHI_{(3)}$ is very sparse, we avoid computing the large and dense matrix $\D^T\Y\S_0$ and multiply only the non-zero entries in $\PHI_{(3)}$. This allows maintaining efficient memory usage and limits the necessary number of CPU cycles. 

\begin{figure}[H]
 \centering
 \includegraphics[width=0.45\textwidth]{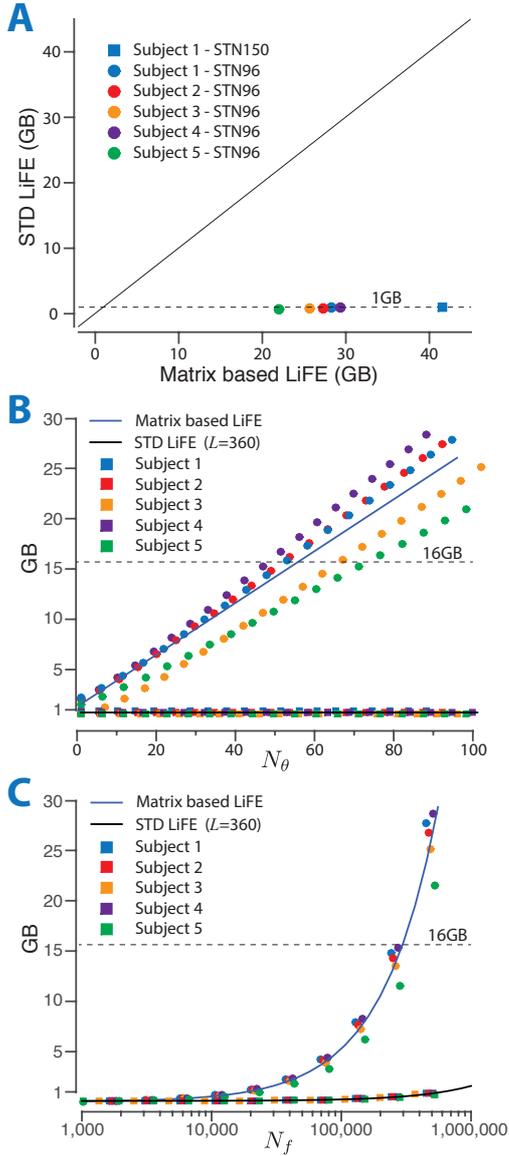}
 \caption{Reduction in memory usage using STD LiFE. \textbf{A}. The memory requirement of $\M$ plotted against memory requirement of the STD model (Eq. \ref{eq:M_Decomposition}, $L=360$, STN96 data, probabilistic tractography, $L_{max}=10$). \textbf{B}. Memory requirement for $\M$ and STD (Eq. \ref{eq:M_Decomposition}) as function of the number of measured diffusion directions ($N_{theta}$, $L=360$, STN96 data, probabilistic tractography, $L_{max}=10$). \textbf{C}. Memory requirement for $\M$ and STD (Eq. \ref{eq:M_Decomposition}) as function of the number of fascicles in the connectome ($N_{f}$, $L=360$, STN96 data, probabilistic tractography, $L_{max}=10$)}
 \label{fig:Fig_sto_Results}       
\end{figure}

\subsection{Model storage reduction}\label{sec:sto_red}

Fig. \ref{fig:Fig_sto_Results}A compares memory usage by $\M$ (Fig. \ref{fig:Fig_LiFE}C; \cite{Pestilli:2014kk}) and the STD LiFE model (Eq. \ref{eq:M_Decomposition}). The storage requirements of $\M$  and the STD model can also be computed analytically. To do so we assume that all fascicles have the same number of nodes $N_n$ and that there are no more than one node per fascicle, per voxel. Under these assumptions the amount of memory necessary to store each fascicle $f$ is proportional to $N_\theta N_n$, thus the storage cost of $\M$ is:
\begin{equation}\label{eq:OLD_LiFE_cost}
C(\M) = \mathcal{O}(N_n N_\theta N_f).
\end{equation}

Conversely, storing fascicles in the STD model require $4N_n$ values only (i.e. the set of the non-zero coefficients and their locations within the core multi-way array $\tPHI$). Thus the amount of memory required by the STD model is:
\begin{equation}\label{eq:NEW_LiFE_cost}
C(\tM) = \mathcal{O}(4N_n N_f + N_\theta N_a),
\end{equation}
where $N_\theta N_a$ is the storage associated with the dictionary matrix $\D\in{\R^{N_\theta \times N_a}}$. Please note that $\S_0$ is absorbed as $\tPHI = \tPHI\times_2 \S_0$ without affecting sparsity, i.e., with no effect on the storage. Storage reduction can be straightforwardly computed as follows:
\begin{equation}
s_{red} = 1 - \frac{4}{N_\theta} - \frac{N_a}{N_n N_f}.
\end{equation}

Fig. \ref{fig:Fig_sto_Results}B shows memory usage for the models in \citep{Pestilli:2014kk} and Eq. \ref{eq:M_Decomposition} as function of number of diffusion directions $N_\theta$ in the data. Given a fixed number of fascicles $N_f$ and nodes $N_n$ in a brain storage of $\M$ grows linearly with the number of diffusion directions ($N_{theta}$; see Eq. \ref{eq:OLD_LiFE_cost}) much faster than the STD model (Eq. \ref{eq:NEW_LiFE_cost}).

Fig. \ref{fig:Fig_sto_Results}C shows memory usage for the models in \citep{Pestilli:2014kk} and Eq. \ref{eq:M_Decomposition} as function of the number of fascicles in a connectome ($N_{f}$). The storage of $\M$ grows linearly with $N_{f}$ and grows much faster than the STD model.

In sum, the STD model provides substantial reduction in memory requirements. The reduction in memory consumption achieved by the STD model becomes more important as the number of measured directions and fascicles in a connectome increase. 

\subsection{Model accuracy}

\begin{figure}[H]
 \centering
 \includegraphics[width=0.45\textwidth]{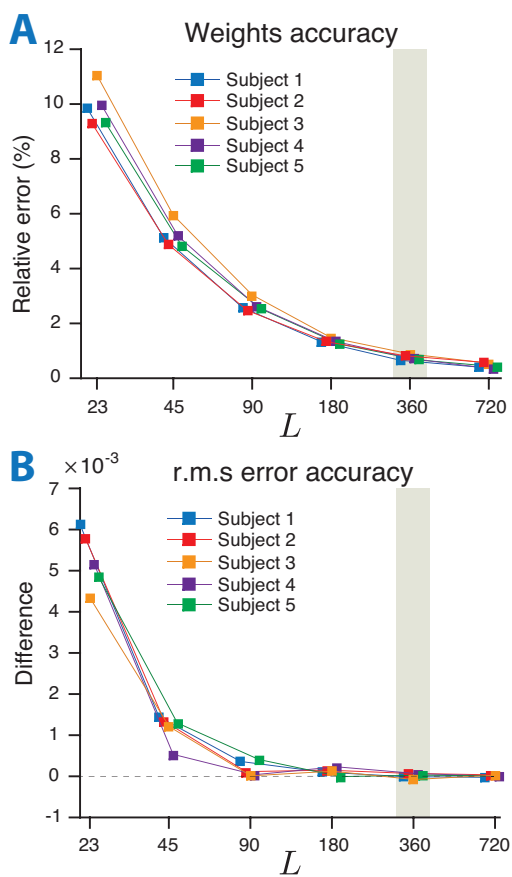}
 \caption{The STD model is as accurate as the original LiFE model. \textbf{A}. Difference in mean cross-validated r.m.s error for a range of discretization steps between the STD and original LiFE models. \textbf{B}. Relative error in estimating the fascicles weights between the STD and original LiFE model for a range of discretization steps (STN96 data, probabilistic tractography, $L_{max}=10$).}
 \label{fig:Fig_accu_Results}
\end{figure}

The STD model provides only an approximation of the original LiFE model. This is because of the discretization introduced by the dictionary ($\D$, see Fig. \ref{fig:Fig_Dictionary}A). Fig. \ref{fig:Fig_accu_Results} shows that the STD model achieves accuracy similar to the original model. We compared the accuracy of the STD model in both, predicting the demeaned signal and estimating the weights of the original model.

Fig. \ref{fig:Fig_accu_Results}A shows the difference between the r.m.s error of the STD and original LiFE model. We fit the STD model using different number of discretization steps of the spherical coordinates (\ref{fig:Fig_Dictionary}A). To do so, the number of discretization steps was varied ($L= 23, 45, 90, 180, 360$ and $720$) and the difference in mean r.m.s. in predicting the demeaned diffusion signal across the whole white-matter volume was computed in reference to the mean r.m.s. of the original model. The mean r.m.s. used for comparison was cross-validated to an independent data set (\citep{Pestilli:2014kk}).

Fig. \ref{fig:Fig_accu_Results}B shows the relative error of the STD model in estimating the weights assigned by the original LiFE model to each fascicle. We fit the STD model using different number of discretization steps of the spherical coordinates as in \ref{fig:Fig_accu_Results}A. We computed the relative error in the estimated weights as $\|\w_0 - \w_1\|/\|\w_0\|$ where $\w_0$, $\w_1$ are the weights of the original and STD models, respectively. Results show that for the data sets tested, $L\ge 360$ allows achieving error below 1\%.

\subsection{Reproduction of results from \citet{Pestilli:2014kk}}
\label{sec:Replicability} 

\begin{figure}[H]
 \centering 
 \includegraphics[width=0.45\textwidth]{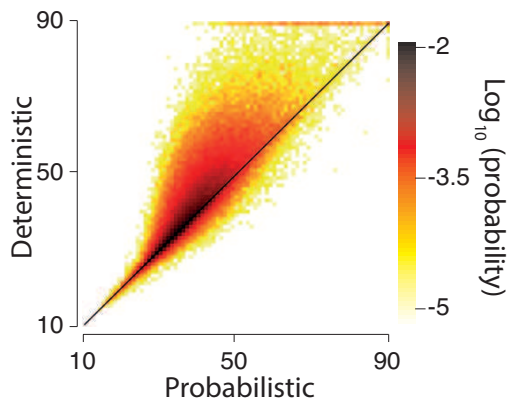}
\caption{The STD model replicates major results of \citep{Pestilli:2014kk}. The scatter plot shows the r.m.s. errors in predicting the demeaned diffusion signal of two tractography models. The r.m.s. error of a probabilistic connectome is plotted against that of a deterministic connectome. R.m.s. was computed using a single subjects from the STN96 data set using CSD-based probabilistic tractography ($L_{max}=10$) and tensor-based deterministic tractography and the STD LiFE model with discretization parameter $L=360$.}
 \label{fig:Fig_replic_Results}
\end{figure}

We demonstrated that a Sparse Tucker Decomposition model achieves similar accuracy to the original LiFE model and reduces storage requirements by 97\%. Fig. \ref{fig:Fig_replic_Results} shows that the STD model replicates a major result of the original \citep{Pestilli:2014kk}. The scatter plot shows the cross-validated  r.m.s. in predicting the demeaned diffusion signal of a probabilistic and deterministic tractography connectome. Results show that the STD model replicates the findings of \citet{Pestilli:2014kk} demonstrating a larger r.m.s. for the deterministic tractogrpahy connectome in a majority of the white matter volume (see Fig. \ref{fig:Fig_Intro}C for the same plot computed using the original LiFE model).

\subsection{Open source LiFE software and reproducibility of results.}
\label{sec:LiFEsoft}
The Matlab implementation of LiFE software using the new STD model is provided at {\small \url{github.com/brain-life/life}} and {\small \url{francopestilli.github.io/life}}.  The Matlab implementation uses the Matlab Tensor Toolbox \citep{tensortoolbox} and mex files compiled for Mac OSX and 64-bit Linux distributions. The software has been tested to run on a standard Notebook computers with less then 8GB of RAM. Processing a whole brain connectome with 500,000 fascicles of the STN150 dataset using a single CPU thread (Notebook computer, 2.2GHz Intel Core i7 processor and 8GB RAM) requires about 5 hours ($L=360$).  

\section{Discussion}
\label{discussion}
The number of brain dataset collected using modern neuroimaging methods is growing at exponentially fast pace. At the same time both data resolution, as well as the size of the populations of human brains being acquired and shared are growing  \citep{VanEssen:2012bo,VanEssen:2013ep,Amunts:2013et, Zuo:2014jr,Calhoun:2011de}. The next generation of brain science will require collaborative efforts between the neuroscience community as well as the informatics and signal processing communities \citep{Garyfallidis:2014ba,Gorgolewski:2011jj,Perez:2007hy}.

The present article focuses on the application of modern signal processing methods for big data to neuroimaging. More speficically we present an example application to diffusion imaging and tractography evaluation \citep{Pestilli:2014kk}. Major value of dMRI and tractography is to allow measuring white-matter in living brains in an individualized manner, one brain at the time. Major efforts are being put forth to improve the representation of the white matter at the macro-, meso- and micro-structural level \citep{Assaf:2008gf,Daducci:fh,Daducci:2015es, Zhang:2012js,Pestilli:2014kk,Assaf:2013fj,Yeatman:2012ku,Yendiki:2011jg,Smith:2006kn,NimmoSmith:2012br,Assaf:2005gf}. The digital nature of neuroimaging data and the availability of large in-vivo databases affords developing new methods to built individual connectomes and compute their accuracy, and validate the results \citep{Pestilli:2014kk}. 

Computational tractography methods exploit the diffusion-weighted signal measured with MRI to identify plausible trajectories of white matter tracts. Most tractography methods estimate candidate tracts one node at a time. Beyond the great excitement brought about by all these technologies, much work is necessary to develop approaches for evaluation and validation of results \citep{Jones:2013gf,2012Sci3371605C,Sporns:2005ki}. To date two primary approaches have been used to tractography validation. First, the geometric accuracy of tractography has been compared to simulated and physical phantoms \citep{2014arXiv1411.5271Z,Fillard:2011fj,Schreiber:2014ca}. Second, tracts and connections identified using tractography have been compared to connections obtained with staining methods in postmortem tissue \citep{Parker:2002kh,Seehaus:2013fc,Azadbakht:2015ea,Thomas:2014ig,Sherbondy:2008ie}. These validation approaches have helped establishing that tractography is accurate to a certain degree and can well identify tracts within the core white matter. 

Attempts to improve tractography have focussed on what is called global tractography. Global tractography methods evaluate the plausibility of the tractography estimates by comparing individual streamlines trajectories with models of the dMRI signal or heuristic rules about streamline smoothness\citep{Jbabdi:2007gp,PierreFillard:2009wy,Li:2012cg,Neher:2012fq,Aganj:2011bc,Fillard:2011fj,Reisert:2011cd,Jbabdi:2008ir,Sherbondy:2008ie}. Alternative approaches to global tractography have proposed the concurrent estimation of the biophysical properties of the white matter fascicles and surrounding tissue \citep{SongZhang:2006uq,Sherbondy:2009wa,Sherbondy:2011wm,Kaden:2015cc,Girard:2015ui}. Historically global tractography methods have been considered computationally intensive tasks with limited applicability to routine study of the human brain in healthy and diseased populations. 

Recently a method for tractography evaluation based on linearized models \citep{Pestilli:2014kk} has been proposed and applied to study living connectomes \citep{Gomez:2015fq,Takemura:2015jr,Yeatman:2014gh}. The method separates the process of tracking from that of global evaluation of the fascicles \citep{Sherbondy:2008ie}. Similarly a family of algorithms based on linearized models have been proposed to solve a series of estimation problems identified in modern tractogrpahy, as well as for studying of the tissue microstructure  \citep{Daducci:2015es,Daducci:fh}. This new generation of linearized methods for microstructure estimation, tractography evaluation are paving the road for new avenues of investigation and the study of the white matter in vivo. 

The contribution of the work presented here is to introduce the general framework of multiway decomposition methods that can potentially be applied to any type of linearized neuroimaging models. The decomposition methods allow reducing the computational complexity of the neuroimaging models and in doing so can increase the impact of these models. The application of multiway decomposition methods to neuroimaging will pave the road to use global estimation and evaluation methods to study large populations of human brains with increasingly high spatial resolution and to apply the modern methods to study the human brain in normal and clinical populations such as those available in modern databases \citep{VanEssen:2012bo,VanEssen:2013ep,Calhoun:2011de}. 

\section*{Appendix A: Computational algorithms.}\label{sec:Algorithms}
Below we report pseudo code for the two operations necessary to fit the decomposed LiFE model. These operations provide the option of being implemented using multi-threading methods. This is because the instances in the loop are independent and allow parallel computations.

\begin{algorithm}[H]
{\footnotesize
\caption{\textbf{: $\mathbf{y}$ = M\_times\_w($\underline{\mathbf{\Phi}}$,$\mathbf{D}$,$\mathbf{S}_0$,$\mathbf{w}$) }} 
\begin{algorithmic}[1]\label{alg:M_times_w}
\REQUIRE Decomposition components ($\underline{\mathbf{\Phi}}$, $\mathbf{D}$, $\mathbf{S}_0$) and vector $\mathbf{w}\in{\mathds{R}^{N_f}}$.
\ENSURE $\mathbf{y}= \mathbf{M}\mathbf{w}$

\STATE $\underline{\mathbf{\Phi}} = \underline{\mathbf{\Phi}} \times_2 \mathbf{S}_0$; \texttt{the result is a very large but still very sparse 3D-array.}
\STATE $\mathbf{Y} = \underline{\mathbf{\Phi}} \times_3 \mathbf{w}^T$; \texttt{the result is a large but very sparse matrix $(N_a \times N_v)$}
\STATE $\mathbf{Y} = \mathbf{D}\mathbf{Y}$; \texttt{the result is a relatively small matrix $(N_{\theta} \times N_v)$}
\STATE $\mathbf{y} = vec(\mathbf{Y})$
\RETURN {$\mathbf{y}$};
\end{algorithmic}
}
\end{algorithm}

\begin{algorithm}[H]
{\footnotesize
\caption{\textbf{: $\w$ = Mtransp\_times\_y($\tPHI$,$\D$,$\S_0$,$\y$) }} 
\begin{algorithmic}[1]\label{alg:Mtransp_times_y}
\REQUIRE Decomposition components ($\tPHI$, $\D$, $\S_0$) and vector $\y\in{\R^{N_{\theta}N_v}}$.
\ENSURE $\w= \M^T\y$
\STATE {$\Y \in{\R^{N_\theta \times N_v}}\leftarrow \y\in{\R^{N_\theta N_v}}$}; \texttt{reshape vector $\y$ into a matrix $\Y$}
\STATE $\underline{\mathbf{\Phi}} = \underline{\mathbf{\Phi}} \times_2 \mathbf{S}_0$; \texttt{the result is a very large but still very sparse 3D-array.}
\STATE {$[\a, \v, \f, \c]$ = get\_nonzero\_entries($\tPHI$)}; $a(n)$, $v(n)$, $f(n)$, $c(n)$ indicate the atom, the  voxel, the fascicle and coefficient associated to node $n$, respectively, with $n=1,2,\dots,N_n$;
\STATE{$\w = \mathbf{0} \in{\R^{N_f}}$}; \texttt{Initialize weights with zeros}
\FOR{$n=1$ \TO $N_n$} 
\STATE{$w(f(n)) = w(f(n)) + \D^T(:,a(n))\Y(:,v(n))c(n)$};
\ENDFOR
\RETURN {$\w$};
\end{algorithmic}
}
\end{algorithm}
%
%

\section*{Acknowledgements}
We thank Hu Chen, Andrzej Cichocki, Aviv Mezer, Ariel Rokem, Richard Shiffrin, Olaf Sporns and Brian Wandell for comments on early versions of the manuscript. We thank Robert Henschel for support using the high performance computers at Indiana University Bloomington and Holger Brunst for support in C code compilation and performance tuning. This project was funded by Indiana University startup funds to F.P.



 
\setlength{\bibsep}{0pt plus 0.3ex}
\small


\end{multicols}

\end{document}